%% file: dsge_crises.tex
\documentclass[reprint,
superscriptaddress,
nofootinbib,
nobibnotes,
 amsmath,amssymb,
 aps,
pre,
]{revtex4-1}

\usepackage{tabularx}
\usepackage{import}
\usepackage{multirow}
\usepackage{comment}
\usepackage[table,xcdraw]{xcolor}
\usepackage{amsfonts}
\usepackage{amssymb}
\usepackage{graphicx}
\usepackage{mathrsfs}
\usepackage{array}
\usepackage[hidelinks]{hyperref}
\usepackage[fleqn,tbtags]{mathtools}
\usepackage{physics}
\usepackage{float}
\usepackage[table]{xcolor}
\usepackage{threeparttable}

\usepackage{tabularx}
\usepackage{appendix}
\usepackage{bbold}
\usepackage{ucs}

\newcommand{\rom}[1]{\uppercase\expandafter{\romannumeral #1\relax}}

\begin{document}

\title{Economic Crises in a Model with Capital Scarcity and Self-Reflexive Confidence}

\author{Federico Guglielmo Morelli}
\email{Both authors contributed equally to this work.}
\affiliation{LPTMC, UMR CNRS 7600, Sorbonne Universit\'e, 75005 Paris, France}
\affiliation{LadHyX, UMR CNRS 7646, Ecole Polytechnique, 91128 Palaiseau, France}
\affiliation{Chair of Econophysics \& Complex Systems, Ecole polytechnique, 91128 Palaiseau, France}

\author{Karl Naumann-Woleske}
\email{Both authors contributed equally to this work.}
\affiliation{LadHyX, UMR CNRS 7646, Ecole Polytechnique, 91128 Palaiseau, France}
\affiliation{Chair of Econophysics \& Complex Systems, Ecole polytechnique, 91128 Palaiseau, France}
\affiliation{New Approaches to Economic Challenges Unit (NAEC), OECD, 75016 Paris, France}

\author{Michael Benzaquen} 
\affiliation{LadHyX, UMR CNRS 7646, Ecole Polytechnique, 91128 Palaiseau, France}
\affiliation{Chair of Econophysics \& Complex Systems, Ecole polytechnique, 91128 Palaiseau, France}
\affiliation{Capital Fund Management, 23-25, Rue de l'Universit\'e 75007 Paris, France}

\author{Marco Tarzia}
\affiliation{LPTMC, UMR CNRS 7600, Sorbonne Universit\'e, 75005 Paris, France}
\affiliation{Institut  Universitaire  de  France,  1  rue  Descartes,  75005  Paris,  France}

\author{Jean-Philippe Bouchaud}
\affiliation{Chair of Econophysics \& Complex Systems, Ecole polytechnique, 91128 Palaiseau, France}
\affiliation{Capital Fund Management, 23-25, Rue de l'Universit\'e 75007 Paris, France}
\affiliation{Acad\'emie des Sciences, Quai de Conti, 75006 Paris, France\medskip}

\date{\today}

\begin{abstract}
In the General Theory, Keynes remarked that the economy's state depends on expectations, and that these expectations can be subject to sudden swings. In this work, we develop a multiple equilibria behavioural business cycle model that can  account for demand or supply collapses due to abrupt drops in consumer confidence, which affect both consumption propensity and investment. We show that, depending on the model parameters, four qualitatively different outcomes can emerge, characterised by the frequency of capital scarcity and/or demand crises. In the absence of policy measures, the duration of such crises can increase by orders of magnitude when parameters are varied, as a result of the ``paradox of thrift''. Our model suggests policy recommendations that prevent the economy from getting trapped in extended stretches of low output, low investment and high unemployment. 
\end{abstract}
\maketitle

\tableofcontents

\textit{The theory can be summed up by saying that, given the psychology of
the public, the level of output and employment as a whole depends on
the amount of investment. I put it in this way, not because this is the
only factor on which aggregate output depends, but because it is usual
in a complex system to regard as the ``causa causans'' that factor which is
most prone to sudden and wide fluctuation. More comprehensively,
aggregate output depends on the propensity to hoard, on the policy of
the monetary authority as it affects the quantity of money, on the state
of confidence concerning the prospective yield of capital assets, on the
propensity to spend and on the social factors which influence the level
of the money wage. 
} 
Keynes \citep{Keynes1936}

\section{Introduction}\label{sec:introduction}

The years following 2008 were marked by the great financial crisis, and with it a crisis for economic theory \citep{Kirman2010}. As for the great depression of the 1930s, there was a failure to predict the crisis amongst economic orthodoxy.\footnote{See the extensive discussions in \citep{Buiter2009, Blanchard2018, DosiRoventini2019, Korinek2017, Romer2016, Stiglitz2018}, with a recent review in \cite{Fair2020}} Despite its failures in predicting the recession \citep{ChristianoEtAl2018} or the sluggish recovery \citep{LindeEtAl2016}, the mainstream Dynamic Stochastic General Equilibrium (DSGE) class of models have remained the core macroeconomic framework and workhorse tool of policy \citep{KaplanViolante2018}. While calls to reform these models have been made \cite{Stiglitz2018, VinesWills2018, VinesWills2020}, the basic framework with a single rational representative agent often remains a baseline assumption when studying business cycles, although heterogeneous agents new Keynesian models (HANK) have recently been considered as well.\footnote{For a non-exhaustive list of prominent examples see \citep{KaplanEtAl2018, KaplanViolante2014, McKayReis2016}.} The decisions of such a representative agent, which include capital investment decisions, determine the trajectory of the economy and are based on the optimisation of a utility function with static parameters, with no space for ``animal spirits'' or confidence effects (although see \cite{BarskySims2012} and the discussion below). In such models, ``dark corners'' are absent and crises can only be the result of major exogenous shocks.\footnote{``Dark corners'' refers to a particularly insightful piece by O. Blanchard in 2014 entitled ``Where Danger Lurks'', see \cite{Blanchard2014}}

This must be contrasted with Keynes' intuition, which, as rephrased by Minsky \citep{Minsky1976}, was that \textit{the subjective evaluation of prospects over a time horizon is the major proximate basis for investment and portfolio decisions, and these subjective estimates are changeable}. Expectations can indeed be subject to rapid changes, disagreement and irrationality, as reflected in the high volatility of investment \citep{StockWatson1999}, and the abrupt nature of expansions and recessions. The investor behaviour behind these swings are indeed often referred to as {animal spirits} or {irrational exuberance} \citep{Shiller2005, AkerlofShiller2010}. There is now a rich literature on irrational behaviour across economics (see \citep{Hommes2021} for a recent review). However, it has not been fully dovetailed into more traditional business cycle models. 
One can find some boundedly rational components in DSGE models,\footnote{
See \citep{FrankeWesterhoff2017} for an early review of animal spirits in macroeconomic models.
} such as \citep{CorneaEtAl2019, HommesLustenhouwer2019, Ozden2021} that focus on learning in expectations formation in a single-actor model, as well as \citep{JumpLevine2019, Gabaix2020} that use various different utility specifications in DSGE. But apart from Refs.  \cite{BarskySims2012,DenizAslanoglu2014,Brenneisen2020}, there is surprisingly little work attempting to factor confidence or sentiment into the DSGE framework as an explicit variable. This is despite some empirical work suggesting that consumer confidence contains important information for forecasting personal spending and consumption \cite{Blanchard1993, CarrollEtAl1994, MatsusakaSbordone1995}. 

Whereas the adapted New Keynesian model of Barsky and Sims \cite{BarskySims2012, Brenneisen2020} conflates confidence with forecasting with imperfect signals, or private news about future technological states, we rather want to focus here on Keynes' animal spirit, self-reflexive facet of confidence, which can be subject lead to abrupt shifts like in 1929 or  2008.\footnote{More recently \cite{AngeletosEtAl2018, AngeletosLaO2013} consider sentiments as uncertainty about the beliefs of others. For another strand of the literature on sudden breakdown of confidence, see \cite{Bouchaud2013,GamaBatistaEtAl2015} and references therein.} As a first step to incorporate such effects and assess their impact on the economy, some of us \citep{MorelliEtAl2020} recently proposed a generalisation of a simple monetary model in which the household's propensity to consume depends on the prior state of the economy, which generates either optimism or anxiety. This feedback can amplify productivity shocks, and lead to the appearance of a second equilibrium characterised by low consumption and high unemployment, and crises resulting from self-induced confidence collapse. The existence of two very different macroeconomic equilibria has also been recently suggested in another context in \citep{CarlinSoskice2018}.   

In the present paper, our aim is to significantly extend the work of \citep{MorelliEtAl2020} by including capital investment as a factor determining the trajectory of the economy. We assume that capital and labour are essentially non-substitutable, and posit a behavioural rule for investment that accounts for both consumer confidence and for the quality of the returns generated by risky capital investment. This expanded framework allows us to investigate the joint dynamics of confidence, capital availability and output. In a nutshell, our model attempts to capture many of the ideas so clearly expressed by Keynes in the opening quote above, while keeping part of the scaffolding of standard business cycle models.  
\begin{figure}[t!]
    \centering
    \includegraphics[width = 0.95\linewidth]{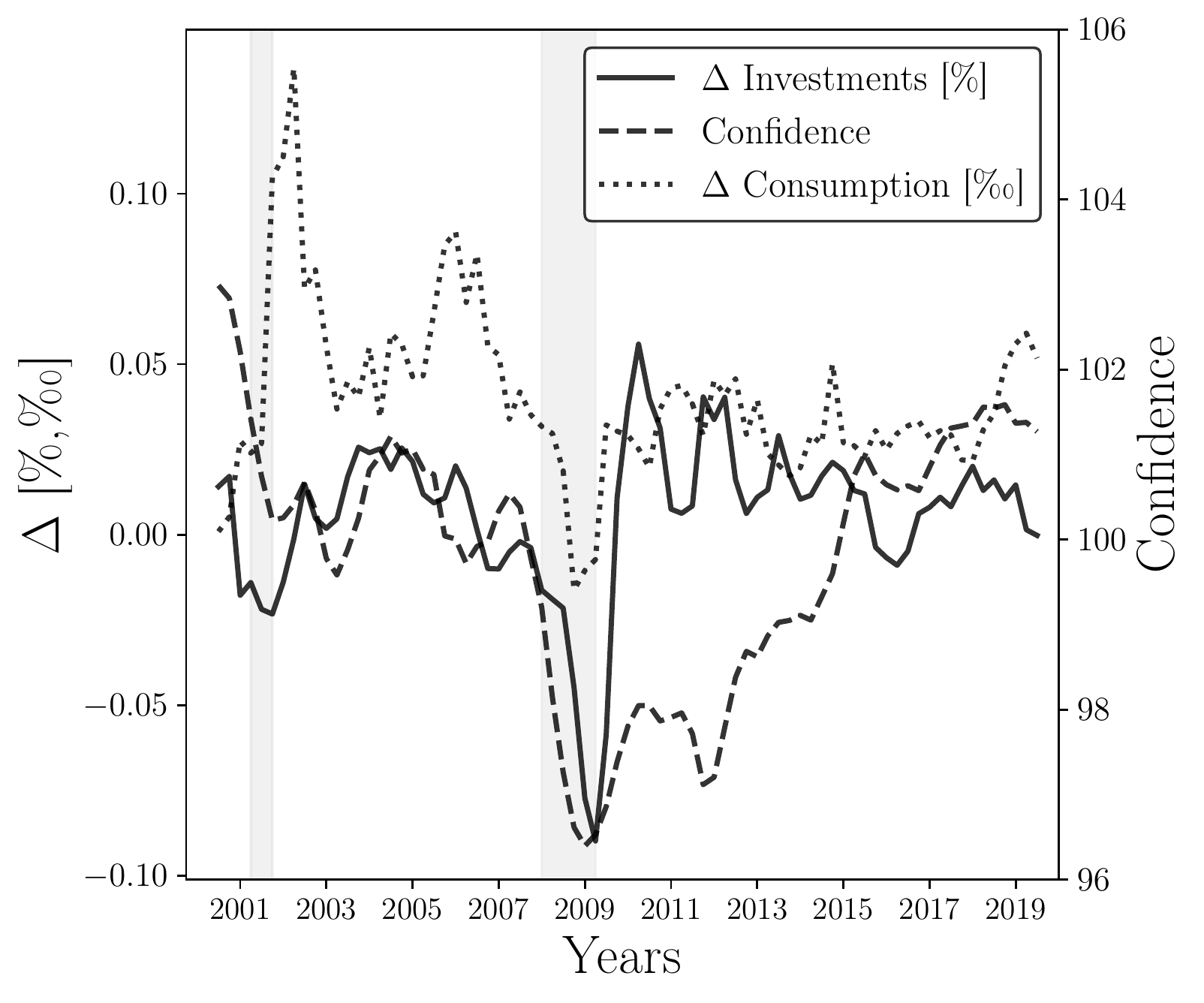}
    \caption{Trajectories of the OECD confidence index, changes in consumption [\textperthousand], and investment [\%] in the United States over a period from the beginning of 2000 to the present. The data were taken from FRED and OECD.
    }
    \label{fig:data}
\end{figure}

In fact, one of our motivations for building such behavioural business cycle models stems from the Great Recession, sparked by Lehman's bankruptcy that led to a sudden collapse in the confidence of both households and investors. This was followed by an almost immediate downfall of both investment and consumption. These stylized facts can be observed in Figure \ref{fig:data}. It took six years to recover to prior confidence levels, even as investment and consumption grew in the medium-term. It is difficult to believe that the Great Recession was the result of a major exogenous shock.\footnote{
\cite{Lo2012} notes that there is no consensus narrative of the causes for the crisis. In addition, from a DSGE model perspective, \cite{AngeletosEtAl2020} recently ruled out many of the common exogenous DSGE shocks as explanatory candidates for business cycles
} Rather, Keynes' story assigning the abruptness of the crisis to a shift in investment decisions is much more plausible. Indeed, anecdotal evidence reported by prominent actors at the time strongly suggests that confidence collapse played an essential role in the unfolding of the crisis -- see the riveting account of Ben Bernanke et al. in \cite{BernankeEtAl2019}.

The consumption and investment trajectories generated by our model can be grouped into four distinct categories differentiated by the prevalence of crises in consumption and the scarcity of capital for production. We recover the bi-stable behaviour obtained in Ref. \cite{MorelliEtAl2020}, alternating between enduring spells of high and low confidence. In addition, the household's investment behaviour can lead to capital scarcity, i.e. periods where capital is the limiting factor to output. During these instances there is an increased risk of a confidence collapse and an ensuing low-consumption depression where the household consumes a small fraction of disposable income and invests cautiously.

Furthermore, when compared with the results of our previous version of the model where capital is absent (or rather, assumed to be so abundant that keeping track of it is unnecessary), we find that low output periods can last orders of magnitude longer. This is because, in the absence of suitable policy measures, investment remains low and capital scarcity prevents the economy from recovering. 

This multiple equilibria scenario is an attempt to move away from the over-simplified, but still dominant single equilibrium paradigm, following recent calls to that effect \cite{VinesWills2020, Greene2021}. Note that the coexistence of different equilibria is also the hallmark of recent agent based models, see e.g. \cite{GualdiEtAl2015,SharmaEtAl2020} and refs. therein.



The qualitative results of our behavioural business cycle model suggests various policy measures, in terms of narratives \cite{Shiller2019} that may change the perception of the future of the economy and the attractiveness of investment in productive capital. In particular, we emphasise the crucial need to maintain capital investment at a sufficiently high level throughout crisis periods, in order to allow for a quick recovery when the economic conditions improve. 

The manuscript is organised as follows. In Section \ref{sec:model} we build up our business cycle model based on \citep{MorelliEtAl2020}, and outline our two novel additions. We then show the various  dynamics the model can generate an reveal its phase diagram in Section \ref{sec:phases}. We discuss in particular how capital scarcity increases the probability of consumption crises, and lead to a multifold increase of the recovery time (see Section \ref{sec:mechanisms_memory}). Section \ref{sec:conclusion} concludes by discussing the policy implications of our findings, and the avenues for possible  extensions of the model. 


\section{A Behavioural Business Cycle Model}\label{sec:model}

The framework presented here hybridises some standard assumptions used in the New Keynesian Dynamic Stochastic General Equilibrium (DSGE) model (see \citep{Gali2015}) with plausible behavioural assumptions about consumption propensity and investment strategies. The environment is based on two blocks: the representative consumer and the representative firm. At this point, we neglect inter-temporal effects and do not attempt to model inflation dynamics and monetary policy, although these features could be included at a later stage. Nonetheless, the phenomenology of our model is already quite rich and needs to be streamlined before exploring further the dynamics of prices. All variables and notations are reported in Table \ref{tab_notation}.    

\subsection{The Household}\label{sec:model_household}

The household sector derives utility from a composite consumption good and provides labour services to the firm. At each time $t$, the representative household maximises its instantaneous utility,
\begin{equation}\label{eq:utility_function}
    U_t(c_t, n_t) :=  G_t \cdot \log c_{t} - \gamma \cdot n^2_{t}\ ,
\end{equation}
where $c_t$ and $n_t$ denote respectively the level of aggregate consumption and the aggregate amount of working hours the household provides to the firm, $G_t$ is the (time dependent) propensity to consume out of income, and $\gamma$ is the disutility of labour (which we fix to 1 for the numerical analysis). 

Each period, the household faces a budget constraint given by its real 
income $\mathcal{I}_t$,
\begin{equation}\label{eq:income}
\mathcal{I}_t := w_t\cdot n_t + \frac{b_{t-1}}{1+\pi_t} + q_{t-1}\cdot \frac{k_{t-1}}{1+\pi_t} \ ,
\end{equation}
which is funded by three sources: (i) the real wage rate $w_t$ paid by the firm for a unit of labour $n_t$, (ii) the real value of the maturating single-period bonds $b_{t-1}$, purchased at time $t-1$ at the price $(1+r_{t-1})^{-1}$ and paying $(1+\pi_t)^{-1}$ at time $t$, where $r_t$ is the interest rate and $\pi_t$ is the inflation rate, and (iii) the realised yield $q_{t-1}$ per unit of real capital $k_t$ that the firm pays to the household in return for investment. We henceforth assume a constant interest rate $r = 0.15\%$ and inflation $\pi = 0.1\%$, keeping in mind a unit time scale corresponding to a month or quarter.  

\begin{figure}[t]
    \centering
    \includegraphics[width= .85\linewidth]{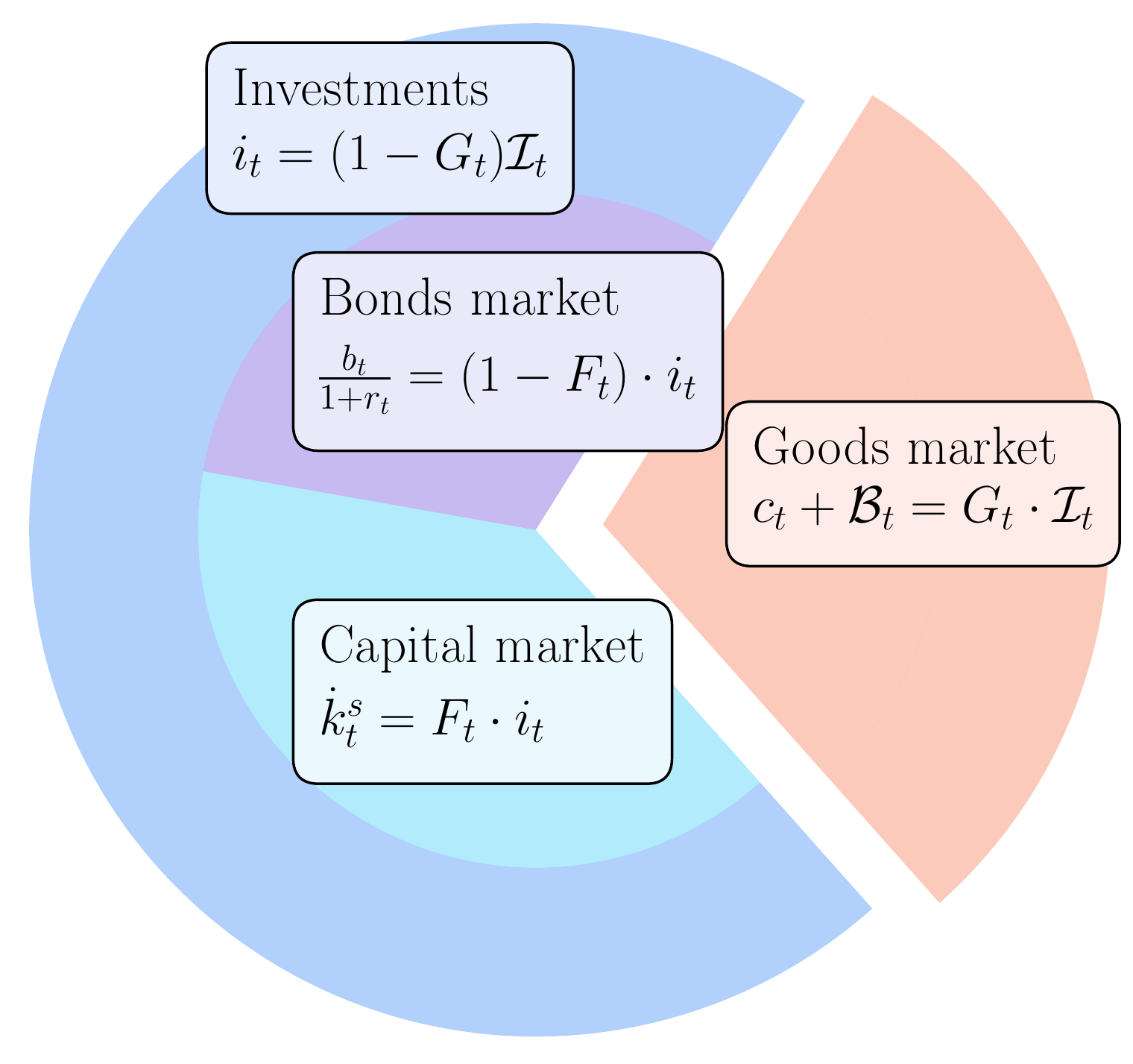}
    \caption{A schematic representation illustrating the division of income, i.e. the budget constraint, by the household.}
    \label{fig:pie_chart}
\end{figure}

Total spending, correspondingly, consists in good consumption (with the price of good set to unity), purchases of new bonds and topping up the firm's capital. Maximisation of the household's utility (Eq. \eqref{eq:utility_function}) leads to the familiar state equation
\begin{equation}\label{eq:state}
 n_t\cdot c_t -\frac{G_t\cdot w_t}{2\gamma} = 0 \ ,
\end{equation}
describing the trade-off between consumption and labour in the current period $t$. 

Interestingly, Eq. \eqref{eq:state} can also be interpreted in a way that lends itself to a natural generalisation for investment decisions. Suppose one starts with a {\it time independent} utility function, Eq. \eqref{eq:utility_function} with $G_t \equiv 1$, which is now optimised under the constraint that the total budget devoted to consumption is a {\it fixed} fraction $G_t \in [0,1]$ of the income $\mathcal{I}_t$, i.e. 
\begin{equation}\label{eq:cons}
c_t = G_t \cdot \mathcal{I}_t . 
\end{equation}
It is easy to show that the very same equation Eq. \eqref{eq:state} immediately follows. We posit that the remaining fraction $1 - G_t$ of income is invested in bonds and capital, i.e. 
\begin{eqnarray}\label{eq:inv}
i_t = (1-G_t) \cdot \mathcal{I}_t \ ,
\end{eqnarray}
where a fraction $F_t \cdot i_t$ (with $F_t\in[0,1]$) is allocated to productive capital, and the remainder 
$(1 -F_t) \cdot i_t$ is invested in bonds -- see Fig.~\ref{fig:pie_chart} for a pie chart summarising the household spending and investment decision.  

The capital level available to the firm thus evolves as
\begin{equation}\label{eq:capital}
    k_{t} = (1-\delta)\cdot k_{t-1} + F_t \cdot (1 -G_t) \cdot \mathcal{I}_t  \ ,
\end{equation}
where $\delta$ is the capital depreciation rate. The remaining investment is allocated to bonds at price $(1+r)^{-1}$, so
\begin{eqnarray}
\frac{b_t}{1+r} &=& (1-F_t)\cdot (1 -G_t) \cdot \mathcal{I}_t  \label{eq:bonds}
\end{eqnarray}
The quantities $G_t$ and $F_t$ aim to capture confidence effects and the attractiveness of risky capital investment, respectively, and are specified in section \ref{sec:model_behaviour} below.

\subsection{The Firm}\label{sec:model_firm}

The economy's productive sector is made up of a single representative firm, which transforms labour $n_t$ and capital $k_t$ into a composite good $y_t$ consumed by the representative household. The firm's production technology is given by a Constant Elasticity of Substitution (CES) function with constant returns to scale,\footnote{In full generality, the CES function should be written as $\left( \alpha\cdot k_t^{-\rho} + (1-\alpha)\cdot (\kappa n_t)^{-\rho}\right)^{-\frac1{\rho}}$, where $\kappa$ is another parameter. However one can always set $\kappa= 1$ at the expense of rescaling the disutility of labour parameter according to $\gamma \to \kappa \gamma$.} 
\begin{equation}\label{eq:prod}
    y_t = z_t\cdot \left( \alpha\cdot k_t^{-\rho} + (1-\alpha)\cdot n_t^{-\rho}\right)^{-\frac1{\rho}} \ ,
\end{equation}
where $\alpha=1/3$ is the capital share in production, $1/(1+\rho)$ is the elasticity of substitution between capital and labour with $\rho>0$, and $z_t>0$ is a stationary exogenous technological process. It is given by $z_t = z_0 e^{\mathfrak{z}_t}$, where $\mathfrak{z}_t$ follows an AR(1) process:
\begin{equation} \label{eq:noise}
    \mathfrak{z}_t = \eta \cdot\mathfrak{z}_{t-1} + \sqrt{1-\eta^2}\cdot\mathcal{N}(0,\sigma^2) \ ,
\end{equation}
with first-order autocorrelation $\eta$, which affects the correlation time of the technology shocks. (In the following we will fix $\eta=0.5$, corresponding to a correlation time of a few months). The base level $z_0$ corresponds to the most probable value of productivity. Note, importantly, that $z_0$ has units of [Time]$^{-1}$, i.e. the amount of goods that can be produced {\it per unit time} for a given level of capital and labour. As our focus is on economic fluctuations, we abstract from production growth in the present model, i.e. the secular dependence of $z_0$ on time.  

The CES production function nests two important limits that affect economic dynamics. As $\rho \to 0^+$, the production function becomes perfectly elastic and recovers the Cobb-Douglas form ({$y^{CD}_t = z_t n_t^{1-\alpha} k_t^{\alpha}$}), whereas in the limit $\rho \to +\infty$ the firm produces via an inelastic Leontief function ({$y^L_t = z_t \min(n_t, k_t)$}).\footnote{Keeping the parameter $\kappa$ free (see previous footnote), the Leontief function would read $y^L_t = z_t \min(\kappa n_t, k_t)$, i.e. $\kappa^{-1}$ measures the amount of labour equivalent to one unit of capital.}  Throughout the following, we choose $\rho = 7$, corresponding to a near Leontief limit, i.e. a very small amount of substitutability between capital and labour. We will briefly comment in section \ref{sec:mechanisms_memory} the impact of higher substitutability.

The firm maximises its target profit $\mathcal{P}_t$ 
\begin{equation}\label{eq:profit}
    \mathcal{P}_t  = p_t \cdot y_t - w_t \cdot n_t - {q}^*_t\cdot k_t \ , \quad (p_t \equiv 1),
\end{equation}
with  respect  to  the labour supply $n_t$ and the capital level $k_t$, where $p_t$ is set to unity and correspondingly $w_t$ is the real wage and ${q}^*_t$ is the real rent on capital. Under the assumption that the market clears, i.e.
\begin{equation}\label{eq:market_clearing}
y_t = c_t \ ,
\end{equation} 
one finds
\begin{eqnarray}
 \tilde{w}_{t} &=& (1-\alpha) \, \left(\frac{\tilde{c}_{t}}{n_t}\right)^{1+\rho}  \label{eq:wage}\\ 
 \tilde{q}^*_{t} &=& \alpha  \, \left(\frac{\tilde{c}_{t}}{k_t}\right)^{1+\rho}  \label{eq:ideal_return}
\end{eqnarray}
where, generically, $\tilde x := x/z$. Note that, as it should be, wages, consumption 
and rent on capital are all in units of $z$, i.e. unit time scale (e.g. a month or a quarter).

Combining the household's state equation, Eq.~\eqref{eq:state} and the equation for the real wage, Eq.~\eqref{eq:wage}, the consumption level $c_t$ must satisfy
\begin{equation}\label{eq:solution}
  \tilde{c}_t^{2} = \frac{G_t}{2\gamma}(1-\alpha)^{\frac2{\rho}} \left[1- \alpha \left(\frac{\tilde{c}_t}{k_t}\right)^{\rho}\right]^{1+\frac{2}{\rho }}\ ,
\end{equation}
As both sides of Eq. \eqref{eq:solution} are monotonous, this ensures a unique solution for any given level of capital $k_t$ and consumption rate $G_t$. As expected, the consumption at time $t$ increases if the capital $k_t$ is increased and/or the consumption rate $G_t$ is increased.


\subsection{The Leontief Limit}
\label{sec:leontief} 
In this section we discuss in detail the Leontief limit of the equations derived in the previous section. Such an analysis will greatly help understanding the dynamics of the model that will be described below. 

\begin{figure}[t!]
    \centering
    \includegraphics[width = .95\linewidth]{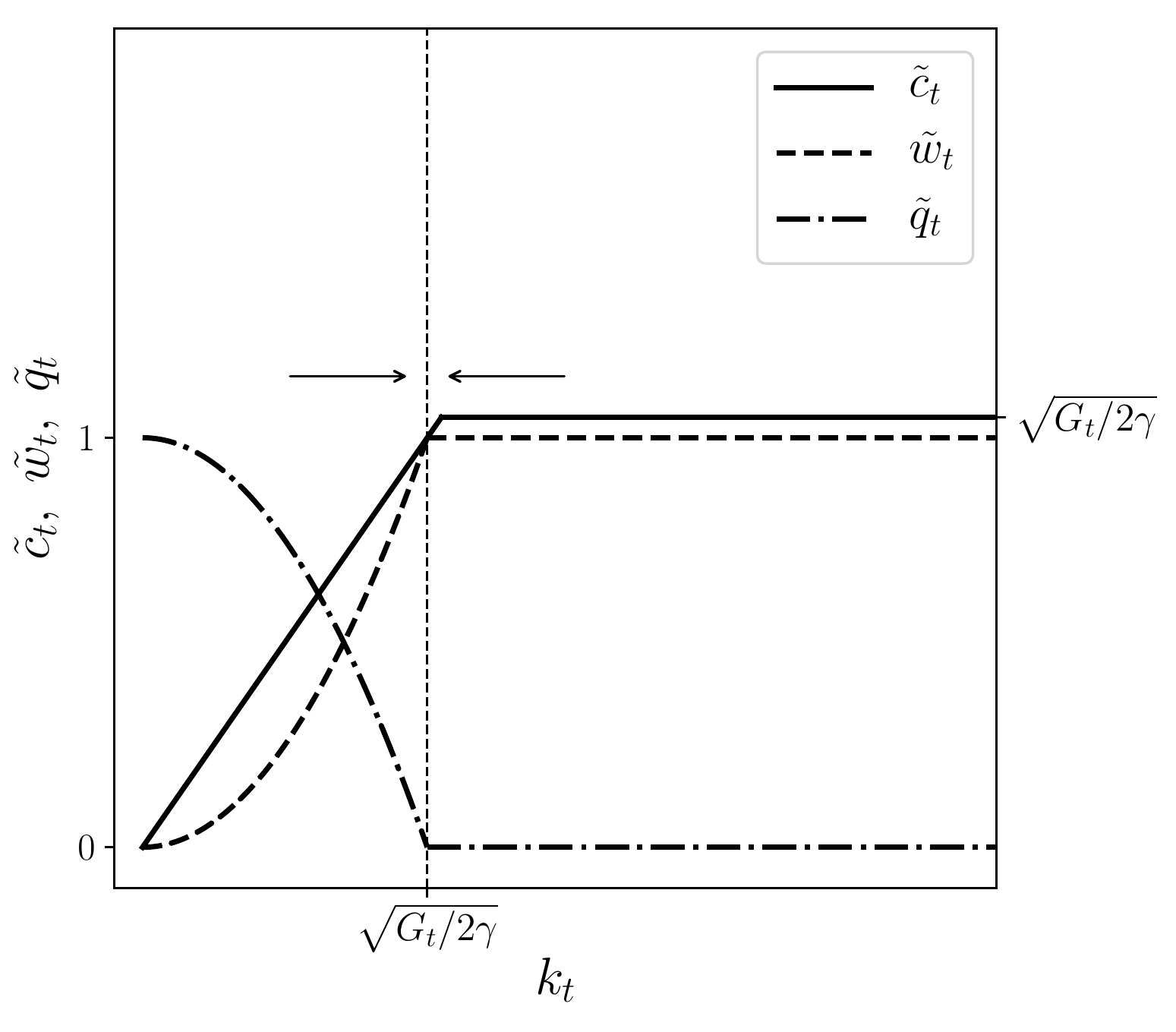}
    \caption{The figure shows the behaviour of rescaled consumption $\tilde{c}_t$, wages $\tilde{w}_t$, and rent on capital $\tilde{q}^*_t$ as a function of $k_t$, in the Leontief limit, i.e. $\rho \to +\infty$. Note the qualitative change of behaviour between  $k_t > \sqrt{G_t/2\gamma}$ and $k_t < \sqrt{G_t/2\gamma}$.
    }
    \label{fig:leontief}
\end{figure}

\subsubsection{Abundant Capital}

Assume first that $\tilde{c}_t < k_t$ and $\rho \to \infty$. Then $(\tilde{c}_t/k_t)^\rho \to 0$ and one finds 
\begin{equation}
\tilde{c}_t \approx \sqrt{\frac{G_t}{2 \gamma}}
\end{equation}
This is only consistent with our working hypothesis when 
\begin{equation}
\frac{G_t}{2 \gamma} < k_t^2.
\end{equation}
In this regime one finds, using Equation \eqref{eq:state}:
\begin{equation} 
n_t = \tilde{c}_t \tilde{w}_t ,
\end{equation}
which once plugged back in Equation \eqref{eq:wage} leads to
\begin{equation}  
\tilde{w}_t = (1 - \alpha)^{1/(2+\rho)} \approx 1.
\end{equation}
Since $ \tilde{c}_t < k_t$, one concludes from Eq. \eqref{eq:ideal_return} that the rent on capital $\tilde{q}_t^*$ is exponentially small. Intuitively, as labour is the limiting factor, consumption is directly proportional to how much the household chooses to work and the productivity at that time, while capital has no distinct effects on the economy.

\subsubsection{Scarce Capital}

Now let us look at the regime 
\begin{equation}
\frac{G_t}{2 \gamma} > k_t^2.
\end{equation}
We introduce the notation $\beta_t := 2 k_t^2 \gamma/G_t$ for further use. We hypothesise that the solution for $\tilde{c}_t$ in this regime is of the form 
\begin{equation}
\tilde{c}_t = k_t e^{-x_t/\rho}
\end{equation}
where $x_t=O(1)$ is to be determined.  Plugging in Eq. \eqref{eq:solution}, we find:
\begin{equation}
k_t^2 e^{-2x_t/\rho} = \frac{G_t}{2 \gamma} (1-\alpha)^{-2/\rho} \left( 1 - \alpha e^{-x_t} \right)^{1 + 2/\rho},
\end{equation}
or, for $\rho \to +\infty$, 
\begin{equation}
e^{-x_t} = \frac{1 - \beta_t}{\alpha}.
\end{equation}
The equation for $\tilde{q}_t^*$ then leads to 
\begin{equation}
\tilde{q}_t^* = \alpha e^{-x_t} = 1 - \beta_t, 
\end{equation} 
which indeed vanishes when $\beta_t = 1$, correctly matching the regime where capital is plentiful, whereas 
$\tilde{q}_t^*$ tends to unity when $\beta_t \to 0$, i.e. where $k_t \to 0$. 

With $\tilde{c}_t = k_t e^{-x_t/\rho}$, one finds from Equation \eqref{eq:state}
\begin{equation} 
n_t = \frac{G_t w_t}{2 \gamma k_t} e^{x_t/\rho}.
\end{equation} 
Finally, plugging into the equation for wages, 
\begin{equation} 
\tilde{w}_t^{2 + \rho} = (1-\alpha) e^{-x_t} \beta_t^{1+\rho},
\end{equation}
or in the limit $\rho \to \infty$, 
\begin{equation}
\tilde{w}_t = \beta_t + O(\rho^{-1}),
\end{equation}
and hence $n_t = k_t + O(\rho^{-1})$. Again, this solution matches with the result $\tilde{w}_t = 1$ obtained for $\beta_t \geq 1$.

\subsubsection{Discussion}

A summary sketch of the above results is provided in Fig. \ref{fig:leontief}, as available capital $k_t$ is varied. Part of the dynamical properties of our model can be inferred from this figure: when capital is lush, return on capital is small and investment decreases (i.e. $F_t$ decreases). If investment falls below the level of capital depreciation $\delta$, then $k_t$ will fall until the level $\sqrt{G_t/2 \gamma}$ is reached. At this point, return on capital $q^*_t$ increases, promoting investment. When consumption propensity $G_t$ increases, $k_t$ may fall behind, again leading to an increase of $q^*_t$. Hence we expect a regime where the economy stabilises close to the point where $k_t \approx \sqrt{G_t/2 \gamma}$, where capital and labour are tracking each other, and interest on capital and wages are neither very small, nor saturated to their maximum value $w_{\max} = q^*_{\max} = z$.

\subsection{The Risk of Investment}\label{sec:model_risk}

Rigidity and costs to capital usage are typically introduced through adjustment costs to capital utilisation (e.g. see \citep{SmetsWouters2007}). In this paper we take a different route. Rather than empowering the household to choose the firm's utilisation rate, we suppose the household invests in capital and gives operational control of capital to the firm. In exchange it is promised a return ${q}^*$ per unit capital and per unit time scale (month or quarter). However, as the volatility of the stock-market attests to, such a return is not assured. Hence, we introduce an intrinsic state-dependent risk to the returns on capital\footnote{
    More sophisticated distributions can be considered. We use this simple form to keep the number of parameters of the model as small as possible.
} $\xi\in[0,1]$ as a modifier, such that the rate actually paid by the firm is:
\begin{equation}\label{eq:real_returns}
    q_t = {q}^*_t\cdot \xi_t \leq {q}^*_t,
\end{equation}
where $\xi$ is distributed as
\begin{equation}\label{eq:real_returns_noise}
    p(\xi) =  a \cdot\xi^{a-1}, 
\end{equation}
where parameter $a$ controls the intensity of the risk. Note indeed that $\mathbb{E}[\xi]=a/(1+a)$ and $\mathbb{V}[\xi]=a/(2+a)(1+a)^2$. Hence the larger the value of $a$, the more $p(\xi)$ is concentrated around $\xi=1$ (full payment). This formulation of risk implies that the representative firm pays out at most the marginal productivity of capital, but more likely only pays a fraction of this, corresponding to an effective description of financial distress and bankruptcy within a representative firm setup. In most simulations we set $a=15$, such that the return is on average $93.75\%$ of the promised return.
In an extended version of the model, the parameter $a$ could itself be a function of the state of the economy (in particular of the availability of capital), but we will not consider this possibility here.

\subsection{Spending and Investing}\label{sec:model_behaviour}

The model laid out in Sections \ref{sec:model_household}-\ref{sec:model_risk} contains two dynamic variables, the consumption rate $G_t$ in Eq. \eqref{eq:cons} and the investment allocation rate $F_t$ in Eq. \eqref{eq:inv}, which have not been specified yet. These two variables are responsible for the feedback mechanisms which are at the core of the dynamical evolution of our model economy.
Here we elaborate on these mechanisms and provide the economic intuition behind them.

\subsubsection{The Consumption Rate} \label{sec:dynamics_consumption}

As in Refs. \cite{MorelliEtAl2020,DenizAslanoglu2014}, we postulate that the consumption rate $G_t$ (or propensity, see section \ref{sec:model_household}) is a function of the {\it consumer confidence index } $\mathcal{C}_t$, that we model as a real variable $\in [-1,1]$ and, possibly, on the difference between the expected inflation rate $\widehat \pi_t := \mathbb{E}_t[\pi]$ and the bond rate $r_t$:
\begin{equation}
    G_t := G_t(\mathcal{C}_t, \widehat \pi_t - r_t, \ldots),
\end{equation}
where the dependence on the second variable is a way to effectively encode the content of the standard Euler equation without explicitly introducing an inter-temporal optimisation of utility, and where the $\ldots$ leaves room to possible additional variables. But since in the present paper we assume both inflation and interest rates to be constant, the second variable will be dropped altogether. 
As far as the first variable is concerned, we follow our previous work in \cite{MorelliEtAl2020}, where we postulated that confidence of a given household is impacted by the level of consumption of {\it other} households in the previous time step. In a mean-field limit, this self-reflexive mechanism writes 
\begin{eqnarray}\label{eq:cons_confidence}
\mathcal{C}_t = \tanh \left(\theta_c\cdot(c_{t-1}-c_0)\right),
\end{eqnarray}
The parameter $c_0$ is a ``confidence threshold'' where the concavity of $\mathcal{C}(c)$ changes (if $c_{t-1}<c_0$, $\mathcal{C}$ is closer to $1$ while if $c_{t-1}>c_0$, $\mathcal{C}(c)$ is closer to $-1$).  Parameter $\theta_c > 0$ sets the width of the consumption interval over which the transition from low confidence to high confidence takes place.
One could introduce, as in e.g. \cite{BeaudryPortier2014, KroujilineEtAl2019, BeaudryEtAl2020}, the impact of macroeconomic news as an extra contribution to the argument of the 
$\tanh$ function. This would describe how the consumer confidence index is further modulated by some exogenous shocks, but we leave such an extension for future work.

Back to the consumption rate $G_t$, we write
\begin{eqnarray}\label{eq:logistic}
     G_t &=& \frac12\big[G_{\min} + G_{\max} + (G_{\max} - G_{\min})\cdot \mathcal{C}_t\big] \,
\end{eqnarray}
where $0\leq G_{\min}<G_{\max}\leq 1$ are the minimum and maximum proportions of income the household will consume. We fix $G_{\min}=0.05$ to ensure the household will consume whenever its income is positive (necessary consumption). Similarly, we set $G_{\max}=0.95$ to account for a minimal form of precautionary savings in response to some uncertainty regarding the future. 

The intuition behind Eqs. \eqref{eq:cons_confidence} and \eqref{eq:logistic} is that when consumption is above the threshold, $c > c_0$, there is high confidence in the future of the economy, thereby a large fraction of income, $G_t \to G_{\max}$, is consumed. High confidence represents the belief that future income will be sufficient to maintain high consumption with a minimal amount of savings to sustain capital levels. Conversely, when consumption is below the threshold, $c<c_0$, the consumption rate collapses, $G_t \to G_{\min}$.  Following a shock or deterioration in consumption to below the confidence threshold, there is uncertainty about the future economy and whether future consumption is assured. This induces the household to save more for the future, effectively reducing the current demand. Economically, $c_0$ is thus similar to minimal consumption: it defines the threshold beyond which there is a panic where the household's ``survival'' is in question.

The parameter $\theta_c$ modulates the households reaction to a breach of necessary consumption, and can be described as the household's panic polarity. For high $\theta_c$, the household requires only a relatively small shock below $c_0$ to reduce the consumption rate to its minimum. This leads to a bi-stable savings behaviour with sharp transitions. Conversely, as $\theta_c\to0$ the household becomes unresponsive to the state of the economy, consuming half its income regardless of high or low preceding consumption. The intermediate levels of $\theta_c$ describe the smoothness of the adjustment to consumption shocks. 

According to \cite{MorelliEtAl2020} there are four distinct ``phases'', i.e. regions of qualitatively comparable dynamics, that are distinguished by the bi-stability of $G_t$. We can observe in particular a phase of high persistent consumption with no crises, high consumption with short downward spikes, or a phase with alternating periods of high consumption (booms) and low consumption (busts).

\subsubsection{The Investment Allocation }

In each period, the household must allocate its savings between one-period bonds and capital. It does so through an allocation decision $F_t$ based on the household's observation of the economy, and its beliefs about future risk and return. The novelty of our model lies in the behavioural foundation that determines the proportion of new investment dedicated to bonds, $F_t$.

Investment beliefs are shaped by two factors: (i) an estimate of the expected risk-adjusted excess returns to capital investment, given by a Sharpe ratio $\mathcal{S}_t$ \cite{Sharpe1966}, and (ii) the current confidence level $\mathcal{C}_t$ about the future state of the economy. 

The Sharpe ratio $\mathcal{S}_t$ is an estimate of the risk-adjusted real return, $q_t-\delta$, of investing capital in the firm versus holding risk-free bonds ($b_t$) paying $r$. It increases as the returns to capital increase or become less volatile. We assume that estimates of the future Sharpe ratio are only based on exponential moving averages of past (observable) realised returns, which is a form of extrapolative beliefs.\footnote{See \cite{DaEtAl2021, KuchlerZafar2019} for recent empirical work on extrapolative beliefs.} In other words, i.e. the mean $\mu^q$ and standard deviation $\sigma^q$ of the return stream are computed as
\begin{eqnarray}
    \mu^q_t &=& \lambda\cdot\mu^q_{t-1} + (1-\lambda)\cdot q_t  \label{eq:ema_mu}\\
    (\sigma_t^q)^2 &=& \lambda\cdot(\sigma^q_{t-1})^2  + (1-\lambda)\cdot (q_t -  \mu^q_t)^2  \label{eq:ema_sigma}\\
    \mathcal{S}_t &:=& \mathcal{N} \cdot\frac{\mu^q_t - r_t - \delta}{\sigma_t^q} \label{eq:sharpe}
\end{eqnarray}
with an exponential moving average defined by a gain parameter $\lambda\in(0,1)$, corresponding to a memory time scale equal to $\mathcal{T}_\lambda := 1/|\log \lambda|$: a larger $\lambda$ implies that a higher weight is given to recent observations. The factor $\mathcal{N} \approx 1/4$ is quite arbitrary, but chosen such that, when compared to the confidence in Eq. \eqref{eq:sentiment_level} below, the two terms are of similar magnitude. (Note that this choice is in fact immaterial, since changing $\mathcal{N}$ is equivalent to rescaling the parameter $\nu$ defined in Eq.~\eqref{eq:sentiment_level} below.)

The interpretation of the Sharpe ratio is as follows: a positive signal $\mathcal{S}_t>0$ suggests that the expected real return to capital investment exceeds the returns to risk-free bonds. The magnitude of $\mathcal{S}_t$ is inversely proportional to the risk of capital investment, as measured by the estimated volatility $\sigma_t^q$. Thus in a high-volatility environment the signal might be positive but weak.

The second indicator potentially influencing the household investment decision is the confidence index, $\mathcal{C}_t$, as previously defined. In periods where the household has low confidence, there is a reduced impetus to invest in risky assets because households wish to guarantee next-period income. These are often periods of crisis with a higher volatility in returns. Since bonds are risk-free, this leads to a higher allocation of funds to bonds, {\it ceteris paribus}. Conversely, higher confidence about the future means more appetite for risk, and hence a higher fraction of the savings invested in the capital of firms and a lower fraction invested in bonds. 

We postulate that the propensity, $F_t$, to make risky bets is a function of the overall \textit{sentiment} $\Sigma_t$, computed as a linear combination of the Sharpe ratio and of the confidence: 
\begin{equation}\label{eq:sentiment_level}
    \Sigma_t = \nu \cdot \mathcal{S}_t + (1-\nu) \cdot \mathcal{C}_t,
\end{equation}
where $\nu\in[0,1]$ is the weight the household gives to its estimates of risk-adjusted return $\mathcal{S}_t$ and its confidence level $\mathcal{C}_t$.
When $\nu=1$, the household's confidence plays no role in the investment rule. For positive Sharpe ratio and confidence indicator, the sentiment is positive, $\Sigma_t>0$, indicating a willingness to invest in risky capital. But if $\nu < 1$ sentiment can turn negative even when the Sharpe ratio is high, because of a high level of anxiety about the future state of the economy, encoded as a negative value of $\mathcal{C}_t$.

Finally, the unbounded sentiment $\Sigma_t$ is transformed into a portfolio allocation to capital $F_t\in[0,1]$ via, 
\begin{equation}\label{eq:allocation}
    F_t = \frac12\left[F_{\max}+F_{\min}+(F_{\max}-F_{\min})\cdot\tanh(\theta_k \cdot\Sigma_t) \right],
\end{equation}
where $F_{\max}$ and $F_{\min}$ represent the maximum and the minimum proportion of total investment $i_t$ invested into capital. In the following, the allocation decision is 
bounded between $F_{\min}=0$ and $F_{\max}=1$, which precludes any divestment (or short-selling) of capital or bonds.\footnote{
    One could allow for divestment by $F_{\min}<0$, however, this would require a more elaborate form for Eq.~\eqref{eq:allocation}.}
The parameter $\theta_k$ represents the sensitivity of the portfolio allocation to the agent's sentiment and sets the width of the sentiment interval over which the capital allocation goes from $F_{\min}$ to $F_{\max}$,  that is how \textit{polar} the investment decision is. For  $\theta_k\to\infty$, the allocation becomes binary, leading to either $F_t= F_{\min}$ when sentiment is negative or $F_t = F_{\max}$ when sentiment is positive.

In the following part, we fix the sensitivities to a rather high value $\theta_k = \theta_c = 15$, such that the transitions between different regimes are sharp.

\subsection{Summary \& Orders of Magnitude}\label{sec:model_dynamics}

In this section we have set up a business cycle model incorporating two behavioural mechanisms: a self-reflexive consumption rate decision, already advocated in Ref. \cite{MorelliEtAl2020}, and an investment allocation decision. The novelty of this paper lies in the behavioural foundation that determines the proportion of new risky investment $F_t$, which depends directly on three key parameters, $\lambda, \nu, \theta_k$, describing the ``sentiment'' of the household, i.e. its risk aversion. $F_t$ also indirectly depends on the risk intensity parameter $a$ and the capital depreciation rate $\delta$. The consumption decision depends on two parameters $c_0$ and $\theta_c$ that define the household confidence about its future welfare.  

In the following we discuss how the parameters of these two feedback mechanisms strongly affect the model's dynamics. Note that a very important parameter of the model is the baseline productivity $z_0$, which fixes the scale of the consumption, wages and rent on capital (all per unit time scale). In the following, we choose $z_0 = 0.05$, corresponding to an annual productivity of capital of $20 \%$ if the unit time step is a quarter and $60\%$ if it is a month.\footnote{
To estimate an appropriate order of magnitude for $z_0$, we considered the gross value added by non-financial corporations in the U.S. divided by the current cost net stock of fixed assets together with total wages (as a proxy for labour), which shows a downward trend to approximately 28\% p.a.
}

Among all the parameters of the model, three have an interpretation in terms of time scales: 
\begin{itemize}
    \item $\eta$, which appears in the dynamics of the productivity shocks, that we have fixed to $0.5$ thoughout this study, corresponding to a time scale $\mathcal{T}_\eta = 1/|\log \eta|$ of a few months ;
    \item $\lambda$, which is the gain parameter used by investors to estimate the Sharpe ratio of risky investments, corresponding to a time scale $\mathcal{T}_\lambda = 1/|\log \lambda|$. Our default value will be $\lambda=0.95$, corresponding to $\mathcal{T}_\lambda \approx 20$ or $5$ years if the unit time is a quarter or a month respectively;
    \item $\delta$, the capital depreciation rate, which we choose in the range $0.001$ -- $0.02$, corresponding to a typical replacement time of capital $\mathcal{T}_\delta = 1/|\log (1-\delta)| \approx 12$ -- $250$ years when the unit time is a quarter, and three times less if it is a month. Hence $\delta=0.001$ means essentially no depreciation of capital.  
\end{itemize}

The role and the effect of varying these timescales is studied in detail in Section \ref{sec:mechanisms_memory}. An important remark, at this stage is that, while our choice of one quarter as the unit time step is quite arbitrary, a combination of parameters that is crucial for the properties of the model is the dimensionless product $z_0 \cdot \mathcal{T}_\delta \approx z_0/\delta$, i.e. how much goods can be produced (per unit capital) over the life-cycle of capital. 
 
\begin{figure*}[htbp!]
    \centering
    \includegraphics[width = .99\textwidth]{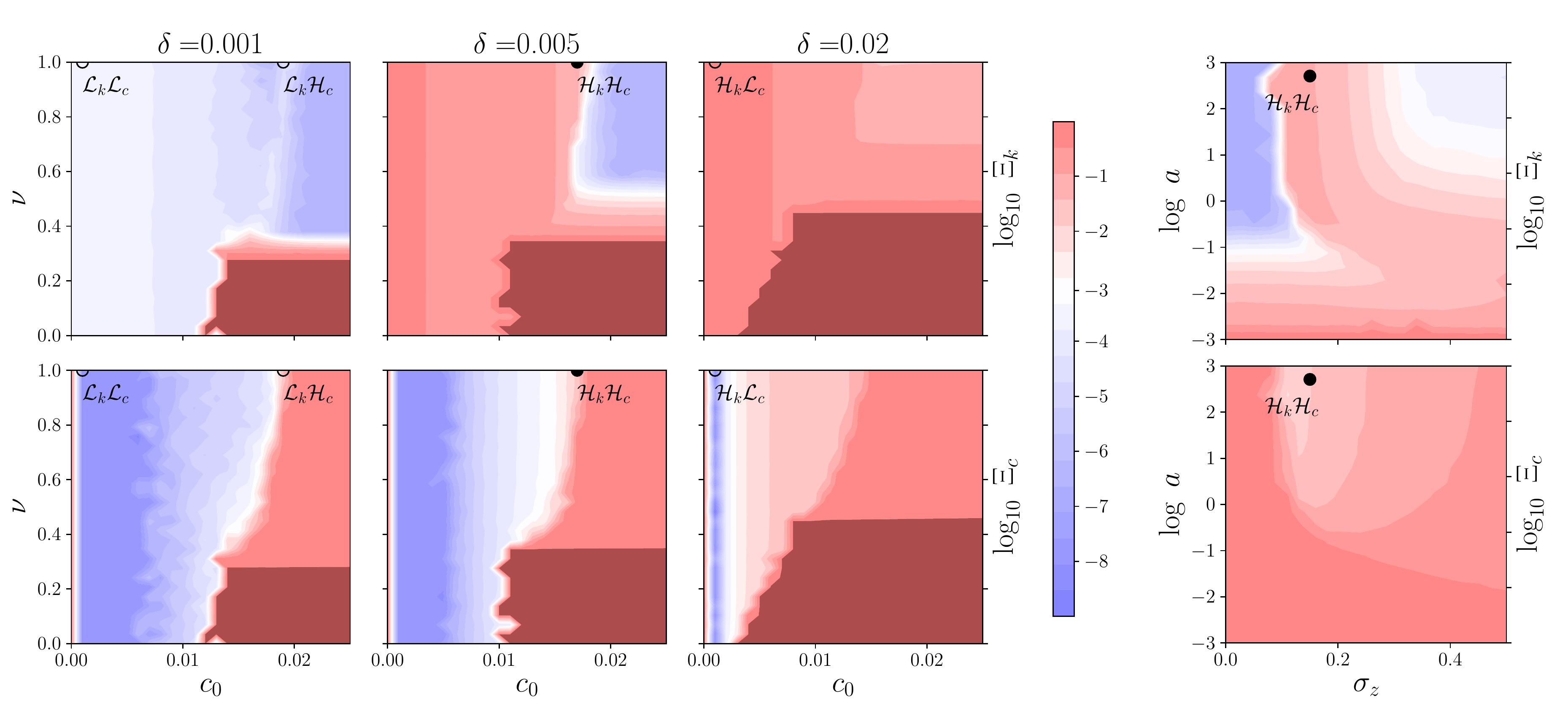}
    \caption{This figure shows different sections of the phase space. The three rightmost upper panels display the $\log_{10}$ probability of capital shortage $\Xi_k$ as a function of the Sharpe ratio weight $\nu$ and of the confidence threshold $c_0$. The bottom row display the $\log_{10}$ probability of consumption crises $\Xi_c$ as a function of the same parameters.
    The dark red zones correspond to the regions where $\Xi_{k/c} > 0.99$, where crises are permanent.
    Each column corresponds to a different choice of the capital depreciation, from $\delta = 0.001$ to $\delta = 0.02$. We set $a = 15$ and $\sigma_z = 0.15$. As $\delta$ increases, the $(\mathcal{H}_k,\mathcal{H}_c)$ regimes becomes widespread.
    In each panel we have marked  
    the points  
    chosen to illustrate  
    the different phases of the model, together with their label (see text). 
    The dynamics trajectories and the corresponding histograms   
    are reported in Fig.~\ref{fig:phase_dynamics}. The two leftmost panels show another section of the phase space, varying $\log\ a$ and $\sigma_z$, with $c_0 = 0.017$, $\nu = 1$ and $\delta  = 0.005$ fixed (the two solid dots there correspond to the same solid dots of the middle panels, 
    in the $\mathcal{H}_k\mathcal{H}_c$ phase).  }
    \label{fig:phase_heatmaps}
\end{figure*}

\section{Crises \& Phase Diagrams}\label{sec:phases}

In this section we first investigate numerically the phase diagram of our self-reflexive business cycle model and highlight the different dynamical features that the model can generate. We choose as control parameters those which govern the behaviour of our two feedback mechanisms: the consumption propensity $G_t$ and the risky investment decision $F_t$. In order to navigate through the following paragraphs, let us explain in a nutshell what is expected to happen in the model.

If a productivity shock causes confidence to drop, consumption propensity $G_t$ and consumption both drop as well, whereas the saving rate $1-G_t$ increases. Because consumption drops, unemployment rises and capital becomes superfluous, leading to a decrease of the rent on capital $q^*$. Because the fraction of savings invested in capital $F_t$ depends both on $q^*$ (through the Sharpe ratio) and on the level of confidence (with a weight $1-\nu$), the amount invested in risky capital, given by $(1-G_t) \cdot F_t \cdot \mathcal{I}_t$ can either increase (if the factor $1-G_t$ dominates) or decrease (if the factor $F_t$ dominates), depending on parameters and conditions. In the second situation, and if capital depreciation is fast, one may face a situation where consumption is impaired and capital becomes scarce at the same time, making recovery more difficult and leading to long periods where the economy is trapped in a low output state. 

\subsection{Crises Indicators}

We focus on two distinct phenomena exhibited by our model: consumption crises and capital scarcity. 
\begin{itemize}
    \item Consumption crises occur in periods where the household's consumption, $c_t$, falls below its threshold, $c_0$. In other words, we have a low demand for consumption which leads the economy into a stagnating low-output state. The severity of such consumption crises is measured as 
\begin{eqnarray}
    \Xi_c &=& \frac1{T}\sum_{t=0}^T \left(1 - \frac{c_t}{c_0} \right) \, \Theta(c_0-c_t),\label{eq:xi_c}
\end{eqnarray} 
where $\Theta(x \geq 0)=1$ and $\Theta(x < 0)=0$ and $T$ is the total simulation time. This indicator counts the fraction of time consumption $c_t$ is low, weighted by the relative distance between $c_t$ and $c_0$. 
    \item Since we are considering an economy defined by low substitutability between capital and labour (i.e. $\rho \gg 1$ in the CES production function), we define capital scarcity as the periods where production is determined by capital levels, i.e. $k_t\leq n_t$. The severity of capital crises is similarly measured as 
\begin{eqnarray}
    \Xi_k &=& \frac1{T}\sum_{t=0}^T \left(1 - \frac{k_t}{n_t} \right) \, \Theta(n_t-k_t).\label{eq:xi_k}
\end{eqnarray} 
\end{itemize}
In a sense, one can consider these two phenomena as demand and supply crises. 
\begin{itemize}
    \item In the consumption crisis state the household does not wish to spend on consumption, hence we see a low aggregate demand. Provided capital depreciation is low, this is also a state of excess capital ($k > n$) and low returns on capital. 
    \item In the capital scarcity state, the firm is bound in its production by the supply of capital, hence it can be viewed as a form of supply crisis. 
\end{itemize}

Both phenomena can be more or less frequent, and at first glance unrelated, but closer scrutiny reveals that in some regions of parameters, these two types of crises interact with one another. To differentiate between characteristic behaviours we distinguish between four different phases in the space of the parameters defined by the values that the indicators $\Xi_k$ and $\Xi_c$ take:
$(\mathcal{L}_k,\mathcal{L}_c)$, $(\mathcal{L}_k,\mathcal{H}_c)$, $(\mathcal{H}_k,\mathcal{L}_c)$, $(\mathcal{H}_k,\mathcal{H}_c)$, 
where $\mathcal{L}$ and $\mathcal{H}$ represent the ``low prevalence'' and ``high prevalence'' of each phenomena $c$ or $k$, respectively. There is however no strict definition of the boundary between high and low prevalence regimes. As a convention, we consider that the crisis prevalence is high when $\Xi \gtrsim 10^{-2}$.

Given this setup, we first focus on the effects of three key parameters: the depreciation rate $\delta$, the weight $\nu$ of the Sharpe ratio in the investment decision, and the consumption threshold $c_0$. Other parameters are fixed to $z_0=0.05$, $\lambda=0.95$ (i.e. $\mathcal{T}_\lambda=20$), $a = 15$ and $\sigma_z = 0.15$.

\begin{figure*}[t!]
    \centering
    \includegraphics[width = .99\textwidth]{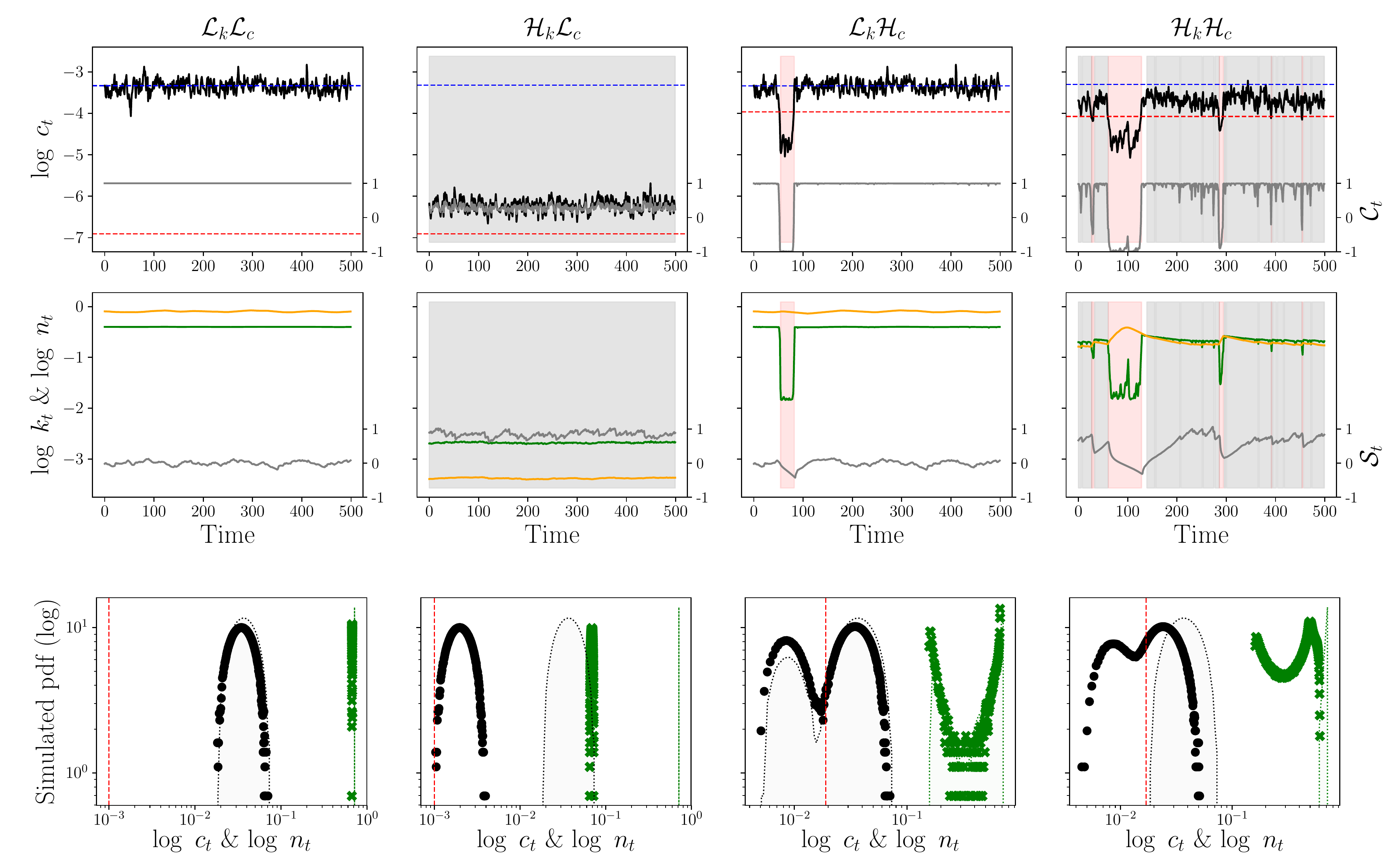}
    \caption{The upper panels show sample dynamics trajectories for each phase presented in Fig. \ref{fig:phase_heatmaps} (marked by circles).  
    In the topmost panels the solid black line corresponds to the consumption $c_t$, while the dashed red and blue horizontal lines show, respectively, the confidence threshold $c_0$ and the average consumption in the $\delta=0$ scenario. Grey (resp. pink) background indicate capital scarcity (resp. consumption crisis).
    The middle row presents the dynamics of capital $k_t$ (solid orange),  labour $n_t$ (solid green) and Sharpe ratio $\mathcal{S}_t$ (solid grey, with levels shown on the right y-axis). 
    The lower panels show the histograms of consumption (black dots) and labour (green crosses) in a $\log-\log$ scale. 
    The green (resp. grey) dashed curve  corresponds to the $\delta=0$ baseline value for labour (resp. consumption), with $c_0$ indicated as a vertical red line.  
    In the $\mathcal{H}_c$ phase, the histograms of consumption and labour become bi-modal, corresponding to high output and low output regimes.
    For all simulations $\nu=1$, $a=15$, $\sigma_z=0.15$. Specific parameters are $\mathcal{L}_k\mathcal{L}_c$: $\delta=0.001$, $c_0=0.001$, $\mathcal{L}_k\mathcal{H}_c$: $\delta=0.001$, $c_0=0.019$, $\mathcal{H}_k\mathcal{L}_c$: $\delta=0.02$, $c_0=0.001$, $\mathcal{H}_k\mathcal{H}_c$: $\delta=0.005$, $c_0=0.017$.}
    \label{fig:phase_dynamics}
\end{figure*}

Figure \ref{fig:phase_heatmaps} presents heat-maps of the severity of capital crises $\log_{10}\Xi_k$ (top) and consumption crises $\log_{10}\Xi_c$ (bottom) across parameter combinations, where red indicates high prevalence. We show two representative sections of the parameter space: the planes $(c_0,\nu)$ (left) and $(\sigma_z,a)$ (right). Since we present the logarithms of $\Xi_{k,c}$, the crossovers between high and low prevalence of different phases are rapid but smooth, i.e. there are no sharp phase transitions that characterise the system's behaviour. From each phase we study a point of the line $\nu=1$ and different values of $c_0$ (marked by points in Figure \ref{fig:phase_heatmaps}), with all other parameters fixed and plot the dynamics of consumption $c_t$, labour $n_t$, capital $k_t$, and the measured Sharpe ratio $\mathcal{S}_t$ in Figure \ref{fig:phase_dynamics}. Note that changing the value of parameters (including $\nu$) while staying in the same phase leads to qualitatively similar trajectories. 

\subsection{Prosperous Stability }\label{sec:phase_stable}

As shown in Figure \ref{fig:phase_dynamics}, leftmost column, the $\mathcal{L}_k\mathcal{L}_c$ phase is characterised by a stable capital surplus, low interest on capital and rare consumption crises. The depreciation of capital $\delta$ is so small that even with a puny level of investment, capital is always in excess and labour is the limiting factor. The stable capital surplus, in combination with a low confidence threshold $c_0$, means that productivity shocks $z_t$ hardly ever reach the required magnitude to trigger a consumption crisis, and if it does, recovery is almost immediate.

The corresponding bottom panel of Figure \ref{fig:phase_dynamics}, shows that the consumption level has normal fluctuations, entirely due to exogenous productivity shocks $z_t$, around a single high-consumption equilibrium. A corollary of the large capital excess is that the labour supply is nearly constant (extremely narrow-distribution in the $\mathcal{L}_k\mathcal{L}_c$ panel of Figure \ref{fig:phase_dynamics}).

As the depreciation rate $\delta$ increases, the average excess of capital supply over labour shrinks, increasing the prevalence of capital scarcity. Accordingly, the phase $\mathcal{L}_k$ quickly disappears upon increasing $\delta$, leading to a pervasive $\mathcal{H}_k\mathcal{L}_c$ phase (see e.g. Fig. \ref{fig:phase_heatmaps}, third column, which shows that capital is always scarce when $\delta=0.02$). As $\delta$ is further increased, the $\mathcal{L}_c$ phase is more and more confined to small values of $c_0$, i.e. when confidence is intrinsically robust. 

\subsection{Prevalent Capital Scarcity }\label{sec:phase_scarcity}

The $\mathcal{H}_k\mathcal{L}_c$ phase is characterised by persistent capital scarcity with rare consumption crises, and is confined within a low $c_0$ ``band'' in the $(c_0,\nu)$ plane when $\delta$ is large enough. Since $c_0$ is low, confidence of is generally high and therefore the household systematically consumes a large proportion $G_t$ of its income, leaving only a small share for investment. Because of capital depreciation, the economy settles in a regime where $k_t < \sqrt{G_t/2\gamma}$, meaning that production is limited by capital, wages are low and rent on capital is high (i.e. the left region in Fig. \ref{fig:leontief}). Hence, the average consumption level is lower than the maximal consumption level reached in the $\mathcal{L}_k\mathcal{L}_c$ phase -- see Figure \ref{fig:phase_dynamics}, second column. 

But since $c_t$ is now closer to $c_0$, consumption crises are lurking around and the economy can flip into the $\mathcal{H}_k\mathcal{H}_c$ if $c_0$ increases and/or if productivity shocks are stronger (higher $\sigma_z$). This is clearly confirmed by the phase diagram of Fig. \ref{fig:phase_heatmaps}. In fact, comparing the phase diagrams for $\delta = 0.005$ and $\delta=0.02$, we see that faster depreciation of capital converts large swaths of $\mathcal{H}_k\mathcal{L}_c$ phase into $\mathcal{H}_k\mathcal{H}_c$. Hence, in this case, investment crises (i.e. the supply side) do trigger consumption crises (i.e. the demand side) by reducing the difference between $c_t$ and $c_0$ -- see also the discussion in section \ref{sec:phase_crises}.    

\subsection{Prevalent Consumption Crisis }\label{sec:phase_cons_crisis}

When the depreciation rate is sufficiently small but the confidence threshold increases, capital remains abundant but self-reflexive confidence crises can hurl the system into a low consumption, low employment regime as a result of random productivity shocks. This is the $\mathcal{L}_k\mathcal{H}_c$ phase. Since capital is high, its level does not impact the level of production, and interest on capital is small. Hence the model becomes completely equivalent, in this regime, to the one studied in \cite{MorelliEtAl2020}, where the dynamics is characterised entirely by the consumption propensity $G_t$ and is dominated by frequent consumption crises, induced by breakdown of collective confidence. 

As argued in \cite{MorelliEtAl2020} and shown in Figure \ref{fig:phase_dynamics}, third column, consumption then displays bi-stable dynamics, where high and low consumption regimes alternate. Correspondingly, the distributions of consumption and labour reveal a secondary peak centred around the low-consumption equilibrium. 


Note that during consumption crises (i.e. $G_t \searrow$) capital becomes even more abundant relative to labour (recall that one needs to compare $k_t$ with $\sqrt{G_t/2\gamma}$) and therefore return on capital and Sharpe ratio both fall, as can be seen in the pink shaded region of Fig. \ref{fig:phase_dynamics}, third column. If we are in a region where consumption crises are short enough compared to both the time $\mathcal{T}_\lambda$ over which the Sharpe ratio is estimated and the capital depreciation time $\mathcal{T}_\delta$, then one can avoid a capital crisis when confidence comes back. Otherwise, the economy enters a turbulent $\mathcal{H}_k\mathcal{H}_c$ phase with both capital and consumption crises.

\subsection{Capital and Consumption Crises}\label{sec:phase_crises}

This final $\mathcal{H}_k\mathcal{H}_c$ phase has both persistent capital scarcity and consumption crises. As anticipated above, capital crises can trigger consumption crises, because capital scarcity drives consumption closer to the confidence threshold $c_0$, below which consumption drops and precautionary savings increase. One can then enter a doom loop (similar to Keynes' famous paradox of thrift) where now capital is too high and leads to a reduction of incentive to invest away from bonds. Hence, as shown in the fourth column of Fig.  \ref{fig:phase_dynamics}, capital and labour fluctuate around low levels, with intertwined periods of capital scarcity (grey regions) and high unemployment (pink regions). The Sharpe ratio gyrates rather strongly between negative values and values close to unity, with a significant negative skewness. The economy is unstable and always far from its optimal state. 

Recall that we have fixed the interest rate on bonds to a constant value. But with a massive demand for bonds, as expected in the $\mathcal{H}_k\mathcal{H}_c$ phase, one should expect the government to borrow at low rates and prop up the economy with public investment, a feature not modelled in the current framework, but certainly worth accounting for in a later version of the model. 

\subsection{Summary}
\label{sec:summary}

To summarise, we have identified four qualitatively different phases of the dynamics. Possibly the most interesting (and novel) one is $\mathcal{H}_k\mathcal{H}_c$, where capital scarcity is persistent, thereby triggering consumption crises.
In this phase the economy is unstable, as capital becomes scarce the likelihood of a consumption crisis increases, and vice versa, low consumption drives rent on capital down and increases the risk aversion of investors. 

We have underlined the role of the capital depreciation rate $\delta$ in determining the fate of our model economy. In fact, when capital and infrastructure are sufficiently long-lived such that $z_0 \cdot \mathcal{T}_\delta$ is large, the economy reaches a stable and prosperous state $\mathcal{L}_k\mathcal{L}_c$, provided self-induced  confidence crises are rare enough (i.e. $c_0$ small). 
Conversely, when $z_0 \cdot \mathcal{T}_\delta$ is low, capital depreciates too quickly and this dents the rents that can be expected by investors. The economy quickly becomes under-capitalised and inefficient, especially because the dearth of capital makes confidence crises more probable, paving the way for the existence of a dysfunctional $\mathcal{H}_k\mathcal{H}_c$ region in the phase diagram.   

\subsection{Investment and Crisis Recovery} \label{sec:mechanisms_memory}

\begin{figure*}[t]
    \centering
    \includegraphics[width = .99\textwidth]{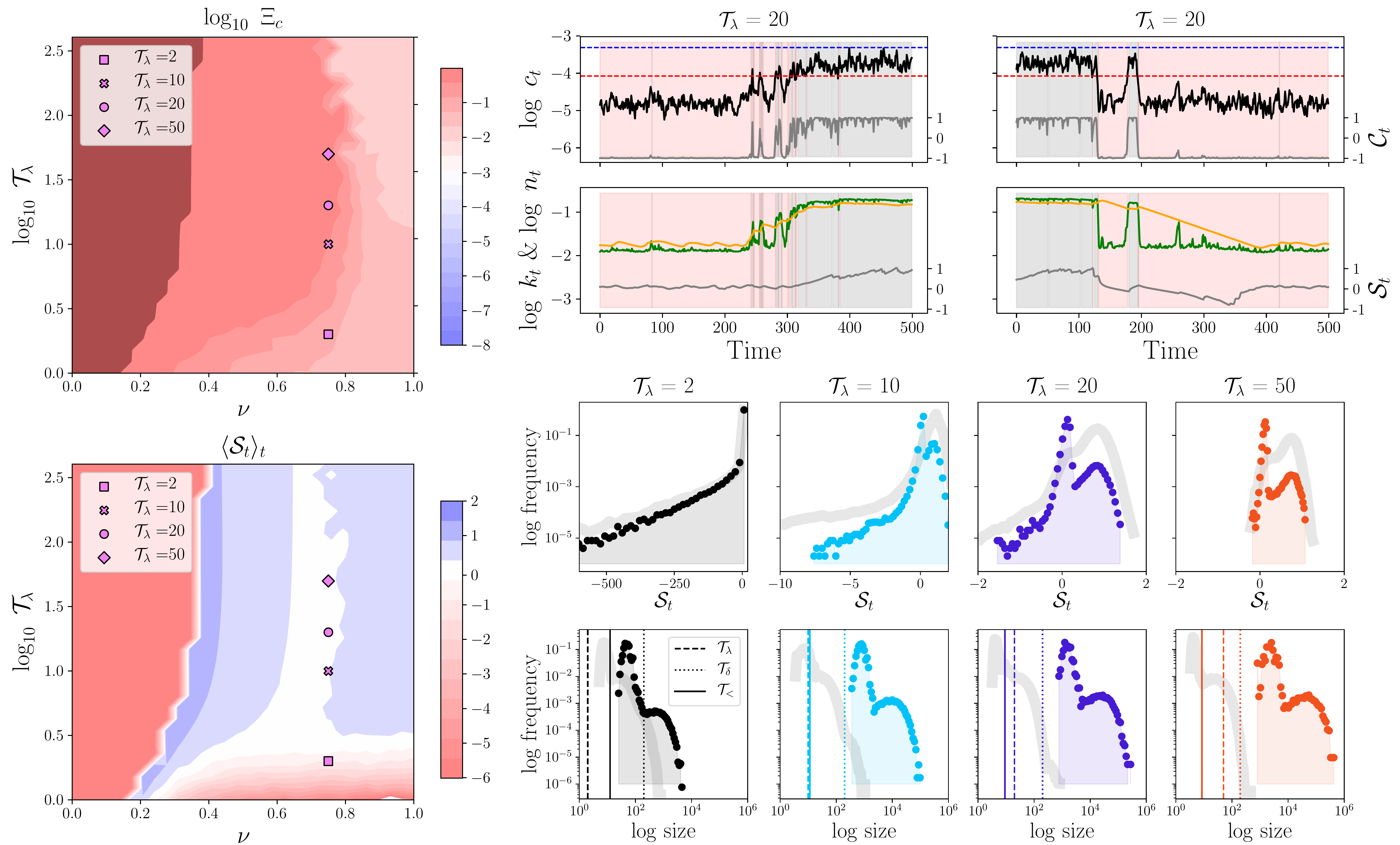}
    \caption{
        The upper (resp. lower) left heat-map shows $\log_{10}\Xi_c$ (resp. average Sharpe ratio) as a function of the memory time scale $\mathcal{T}_{\lambda}$ and the weight parameter $\nu$. The dark red zones correspond to the regions where $\Xi_c > 0.99$, where crises are permanent. 
        The right set of panels shows snapshots of the dynamics of $c_t$, $k_t$ and $n_t$, corresponding to recovery (left) and crisis formation (right), both for $\mathcal{T}_{\lambda}=20$ and $\nu=0.75$. The two bottom rows show the histograms of Sharpe ratio $\mathcal{S}$ and crisis duration $\mathcal{T}_{<}$ for four values of $\mathcal{T}_{\lambda}$: 2, 10, 20 and 50, all for the same value of $\nu$ (shown as symbols in the two heat-maps on the left). The faded grey lines show the same histograms in the benchmark case $\nu=1$ that correspond to the $\mathcal{H}_k\mathcal{H}_c$ point in Fig. \ref{fig:phase_heatmaps}. Other parameters used are: $\delta=0.005$, $c_0 = 0.017$, $\sigma_z = 0.15$ and $a = 15$.
}
    \label{fig:mechanisms_memory_phase}
\end{figure*}


In the previous section, we have explained how capital depreciation can cause instabilities, and the appearance of a $\mathcal{H}_k\mathcal{H}_c$ phase where both capital and consumption undergo regular crises. In this section, we want to explore the influence of the memory timescale $\mathcal{T}_\lambda$, which is the history span over which investors assess the Sharpe ratio of capital investment, and of the sentiment parameter $\nu$ on the time needed for recovery when in a crisis period. We focus on this turbulent phase of the economy. We will fix the other two timescales $\mathcal{T}_{\delta}, \mathcal{T}_\eta$ defined in section \ref{sec:model_dynamics} to, respectively, $200$ and $2$.

Our benchmark will thus be the $\mathcal{H}_k\mathcal{H}_c$ point in Fig. \ref{fig:phase_heatmaps}, corresponding to 
$\lambda=0.95$, $\delta=0.005$, $\nu=1$ and $c_0=0.017$ (with $a$ and $\sigma_z$ also fixed at their baseline values). Looking at the statistics of the high consumption periods and of the low consumption periods, we conclude that the prosperous periods last a time $\mathcal{T}_{>}$ of the order of $\mathcal{T}_{\delta}=200$ (data not shown), whereas crises are rather short, of the order of $\mathcal{T}_{<} \approx 10$, see Fig. \ref{fig:mechanisms_memory_phase}, plain vertical lines in the third graph of the bottom row, which corresponds to 
$\mathcal{T}_\lambda = 20$, i.e. $\lambda=0.95$. The full distribution of $\mathcal{T}_{<}$ and of the Sharpe ratio $\mathcal{S}$ for $\nu=1$ are shown in light grey, and reveals that whereas its time averaged value of $\mathcal{S}$ is clearly positive and equal to $\langle \mathcal{S} \rangle_t \approx 0.71$, its full distribution is uni-modal but quite broad and negatively skewed.

Fig. \ref{fig:mechanisms_memory_phase}, left graphs shows the consumption crisis prevalence $\Xi_c$ and the average Sharpe ratio $\mathcal{S}$ as $1-\nu$ (weighing confidence in the investment allocation decision) and $\mathcal{T}_\lambda$ are varied. One sees that decreasing $\nu$ or increasing  $\mathcal{T}_\lambda$ leads to an increase of $\Xi_c$, at least in the range shown,  $\mathcal{T}_\lambda \lesssim 300$. The evolution of the average Sharpe ratio is more complex, reflecting the non-trivial shape of its distribution function (bi-modal and skewed, see below). But certainly as agents pay less attention to the actual return on capital and are more affected by the level of confidence, the average Sharpe becomes strongly negative (red region of the diagram) and the economy gets trapped forever in a low consumption, low investment regime where $\theta_k \Sigma$ is negative (see Eq. \eqref{eq:allocation}). 

Now, let us look at a cut along the direction $\nu=0.75$, corresponding to a $25 \%$ weight given to confidence in the allocation decision, as $\mathcal{T}_\lambda$ is varied. For this particular value of $\nu$, the average Sharpe ratio is close to zero and only weakly depends on $\mathcal{T}_\lambda$ (Fig. \ref{fig:mechanisms_memory_phase}, bottom left graph). But from the bottom row of Fig. \ref{fig:mechanisms_memory_phase}, we see that 
when $\mathcal{T}_\lambda \gtrsim 10$, the distribution of Sharpe ratios becomes {\it bi-modal} and with a skewness that decreases as $\mathcal{T}_\lambda$ increases. This can be rationalised as follows:

\begin{itemize}
    \item The peak corresponding to positive Sharpe ratios comes from prosperous periods, where consumption is high and capital relatively scarce, leading to a positive return on capital $q^*$: see Fig. \ref{fig:mechanisms_memory_phase}, top right graphs: in the high consumption phase, the orange line (capital) is below the green line (labour). 
    \item The peak corresponding to zero Sharpe comes from crises periods, where capital is in slight excess of labour, leading to a small return on capital (see again Fig. \ref{fig:mechanisms_memory_phase}, top right graphs, and section \ref{sec:leontief}). 
    \item The fat left tail corresponding to negative Sharpes comes from the transitory periods between high confidence and low confidence, when consumption and labour collapse but capital depreciates much more slowly. In this case, return on capital plummets and the Sharpe ratio becomes negative.
    \item As $\mathcal{T}_\lambda$ increases, the weight of these transitory regimes in the estimate of the Sharpe ratio becomes small, and the fat left tail disappears, as crises become less frequent and much longer.\footnote{For very large $\mathcal{T}_\lambda$, the situation changes again, see below.}
\end{itemize}

Whereas the length of the prosperous periods $\mathcal{T}_>$ is unchanged compared to the benchmark $\nu=1$ case for all values of $\mathcal{T}_\lambda  \gtrsim 10$, the length of the crisis periods $\mathcal{T}_<$ increases by more than a 100 times as $\nu$ is decreased from $1$ to $0.75$ (compare the grey line and the coloured points in bottom row in Fig. \ref{fig:mechanisms_memory_phase}). The first observation is due to the fact that the Sharpe ratio estimated when in a high output period is clearly in positive territory and quite insensitive to $\mathcal{T}_\lambda$ (see the histograms in Fig. \ref{fig:mechanisms_memory_phase}). This means that capital supply is also independent of $\mathcal{T}_\lambda$ and that the confidence collapse mechanism must be identical to the one described in our previous work \cite{MorelliEtAl2020} and not triggered by a lack of investment.\footnote{
This is not to say that the crisis frequency is not related to capital abundance. As already noted in section \ref{sec:phase_crises}, as capital depreciation $\delta$ increases, available capital decreases, which leads to a lowering of output $c_t$. Since the distance between $c_t$ and the threshold $c_0$ is a crucial determinant of the probability of a confidence crisis, the region $\mathcal{H}_k\mathcal{H}_c$ becomes pervasive as $\delta$ increases, see again Fig. \ref{fig:phase_heatmaps}.
}

On the contrary, the mechanism by which confidence is {\it restored} is strongly impacted by the value of the memory time $\mathcal{T}_\lambda$. When the economy is in a consumption crisis, the returns on capital are very small. Thus, averaged over a sufficiently long time period, the Sharpe ratio is well defined and also small (see the narrow peaks in the Sharpe ratio distribution in Figure \ref{fig:mechanisms_memory_phase}). Combined with the low confidence dampener on sentiment $(1-\nu) \cdot \mathcal{C}_t$, this leads to a negligible investment flow. So whereas positive productivity shocks should put the economy back on an even keel, the level of capital is lagging, which creates a ceiling that prevents consumption (and hence confidence) from increasing substantially and returning to the high consumption case. 

Interestingly, the dependence of $\mathcal{T}_<$ on $\mathcal{T}_\lambda$ is in fact non-monotonic. For very large $\mathcal{T}_\lambda \gtrsim 1000$, the memory of prosperous periods persists even during the crises, so that the Sharpe ratio and investment always remain high. In such cases, $\mathcal{T}_<$ abruptly drops back to small values $\lesssim 5$ (data not shown). With extremely small probability, however, the system remains trapped in a crisis forever.

In the opposite case of a small enough  $\mathcal{T}_\lambda$, the short periods where consumption increases due to productivity shocks allow sufficiently rapid increases in capital rent to encourage immediate investment. This is enough to prop up capital and allows confidence to be fully restored as labour and consumption will grow with the limiting factor $k_t$. For an example of these positive spikes of consumption, see top centre panel in Fig. \ref{fig:mechanisms_memory_phase}. The same effect takes place if $\nu$ is increased back to $1$, where only realised Sharpe affects investment. In this case, the drag on capital due to low confidence levels is absent, and the system is able to pull itself out of the rut much more efficiently, leading to shorter crisis periods. But for lower values of $\nu$ (higher impact of household confidence on the investment propensity), the dearth of capital in crisis periods is such that the economy is unable to ever recover, i.e. $\mathcal{T}_< = + \infty$ for all purposes. 

From a policy point of view, reducing interest rates has the direct effect of increasing the Sharpe ratio and reducing the return to bonds, thus promoting investment and making the transition back to the high consumption state easier. However, this may 
require the central bank to set interest rates $r$ to negative values, as $r$ which might already be close to zero due to prior crises. Besides monetary policy, other measures that improve confidence (e.g. central bank messaging) and/or promote investment into productive capital would have a similar impact (for instance if the government decides on strong fiscal measures that include investment into productive capital, such as through mission-oriented policies or infrastructure spending).

Finally, let us mention that while the existence of consumption crisis is independent of the substitutability parameter $\rho$, the duration of the low investment, low consumption periods is also highly sensitive to substitutability effects. We have indeed found that when $\rho$ is sufficiently small, i.e. for production functions closer to Cobb-Douglas than to Leontief, recovery is much faster (data not shown). This could have been expected: lack of capital can now be compensated by labour, expediting the transition back to a prosperous state of affairs.   

\section{Discussion and Conclusions}
\label{sec:conclusion}

We have constructed a behavioural real business cycle model in which labour and capital are nearly unsubstitutable. In the model, consumption and investment are controlled by (a) the confidence of households, which is self-reflexive (i.e. agents take cues from the consumption of other agents to determine their consumption budget) and (b) the quality of the excess real return to capital, as measured by the Sharpe ratio. As we have shown in our previous work \cite{MorelliEtAl2020}, the self-referential nature of confidence amplifies the effect of productivity shocks on output, and can lead to crises where consumption abruptly jumps from a high equilibrium level to a low equilibrium level. Depending on the parameters of the model, these crises can be more or less frequent, and the low consumption periods can be of various duration: short spikes (``V-shaped crises'') or long drawn-out phases (``L-shaped crises''). 

In the present study we investigate how the introduction of capital affects these dynamical patterns. In our model, capital can either be abundant (in which case labour is the limiting factor to production) or scarce. The main factors determining the quantity of working capital are the depreciation rate and the propensity of the households to save and invest, which itself depends on the return on capital. The resulting phenomenology of the model is quite rich. Our analysis reveals the following main takeaways:
\begin{enumerate}
    \item Higher capital depreciation rates, {\it ceteris paribus}, lead to capital scarcity and limit production. This makes the economy more prone to confidence crises, increasing their prevalence;
    \item Increasing the influence of the level of confidence in capital allocation decisions creates a feedback loop similar to Keynes' paradox of thrift, destabilising and trapping the economy into a non-optimal low consumption state;
    \item The time during which the economy remains in a low output state is highly sensitive to the time span  over which investors compute the Sharpe ratio.  Increasing this memory timescale leads to sluggish adjustments of investment.\footnote{Note however, as reported in the previous section, that extremely long memory timescales allow the Sharpe ratio to stick to high values.} Consequently, instantaneous increase of capital returns due to productivity upticks are not sufficient to boost the investment propensity. This leads to a persistence of capital scarcity, and prevents the economy from escaping the low output trap.
\end{enumerate}

Our findings have different policy implications. As already emphasised in our previous paper \cite{MorelliEtAl2020}, if self-reflexive feedback loops exist, then governments and monetary authorities should not only manage inflation expectations but more broadly confidence in the future prospects of the economy. Although confidence indices are routinely measured by polling institutes (see e.g. Fig. \ref{fig:data} in Section \ref{sec:introduction}), the inclusion of such indices in macroeconomic DSGE models and the importance of narratives \cite{Shiller2019} have never really been considered seriously beyond the impact of news shocks on productivity.\footnote{
e.g. the work of \cite{BeaudryEtAl2020} reflecting interactions and complementarity. Also \cite{AngeletosEtAl2020} showed that single news shocks (confidence shocks) are sufficient to fit empirically the effects of business cycles. See also \cite{DenizAslanoglu2014}.
}

Beyond communication and narratives, our model suggests that monetary authorities should also directly promote investment, in particular during recessions. This is needed to prevent the economy being trapped in a low output, low confidence environment. Although this conclusion looks perfectly intuitive, our model reveals that a lack of capital can prolong crisis periods by orders of magnitude, and convert V-shaped crises into L-shaped crises. Boosting investment in working capital can be done through traditional channels, by lowering the risk-free interest rate (possibly making it negative) or by direct Keynesian investments in infrastructure and in innovation, which have the double effect of increasing the productivity of capital and propping up household confidence.  

There are of course many directions in which our model should be extended and improved. The first obvious one is to allow interest rates and inflation to be dynamical variables, and to introduce an explicit monetary policy with the central bank monitoring inflation and confidence. A fully developed DSGE model building upon the framework proposed here would be welcome. Other relevant extensions could be to include a feedback mechanism between confidence and the time scale $\mathcal{T}_\lambda$ or the sentiment parameter $\nu$. This would allow potentially relevant panic effects to set in the model, and capture what happened in 2008, for example. 

Another possible extension is to allow the parameter $a$ which describes the default risk on capital to depend on the state of the economy, since bankruptcies are more frequent when the economy is in a low output, low investment regime. 

Last, but not least, we have assumed that confidence is only a function of past realised output, but other factors should obviously be taken into account to model the dynamics of the confidence index, in particular financial news (like in 2008) or geopolitical news. Our framework would lead to scenarios where a shock like Lehman's bankruptcy simultaneously affects both consumption and investment, leading to a deep and prolonged recession, even in the absence of any ``true'' productivity shock. Conversely, good news about the future (e.g. technology shocks) could help recovering faster from the low output trap.

It would also be interesting to look for a ``grand unification'' between the type of behavioural business cycle/DSGE models considered in this paper and heterogeneous agent based models studied in the recent literature, which generically give rise to similar crises and bi-stable dynamics between high output and low output regimes of the economy (see e.g. \cite{GualdiEtAl2015, SharmaEtAl2020}).
\vskip .2cm\noindent
\paragraph*{Acknowledgements}
We deeply thank Davide Luzzati who contributed to the early stages of this work, and Francesco Zamponi for many insightful discussions.  
This research was conducted within the Econophysics \& Complex Systems Research Chair, under the aegis of the Fondation du Risque, the Fondation de l’Ecole polytechnique, the Ecole polytechnique and Capital Fund Management. Karl Naumann-Woleske also acknowledges the support from the New Approaches to Economic Challenges Unit at the Organization for Economic Cooperation and Development (OECD).

\vskip .2cm\noindent
\bibliographystyle{unsrt}
\bibliography{Paper_DSGE.bib}

\newpage
\appendix
\input{tab_notations}{}

\newpage

\end{document}

%% file: tab_notations.tex
\begin{table*}[t]
\centering
\caption{Notation used in the model} \label{tab_notation}
\begin{ruledtabular}
\begin{tabular}{lllr}
{} & {\textbf{Description}} & {\textbf{Definition}} & {\textbf{Value(s)}}\\
\multicolumn{4}{l}{\textbf{Variables}}\\
$c_t$           & Real consumption                  & Sec. \ref{sec:model_household} Eq. \eqref{eq:utility_function} & \\
$n_t$           & Labour                            & Sec. \ref{sec:model_household} Eq. \eqref{eq:utility_function} & \\
${G}_t$         & Propensity to consume from Income & Sec. \ref{sec:model_household} Eq. \eqref{eq:logistic} & $[G_{\min}, G_{\max}]$ \\
$\mathcal{I}_t$ & Income in period $t$              & Sec. \ref{sec:model_household} Eq. \eqref{eq:income} & \\
$w_t$           & Real wage                         & Sec. \ref{sec:model_household} Eq. \eqref{eq:wage} & \\
$b_t$           & Real income from one-period bond purchased in $t$       & Sec. \ref{sec:model_household} Eq. \eqref{eq:bonds} & \\
$k_t$           & Real capital                      & Sec. \ref{sec:model_household} Eq. \eqref{eq:capital} & \\
$q_t$           & Realised yield on capital         & Sec. \ref{sec:model_household} Eq. \eqref{eq:capital} & $[0, {q}^\star_t]$\\\
$i_t$           & Real value of total investment    & Sec. \ref{sec:model_household} Eq. \eqref{eq:inv} & $(1-G_t)\mathcal{I}_t$ \\
$F_t$           & Allocation of investment to capital & Sec. \ref{sec:model_household} Eq. \eqref{eq:allocation} & $[F_{\min}, F_{\max}]$ \\
$z_t$           & Total factor productivity (TFP)   & Sec. \ref{sec:model_firm} Eq. \eqref{eq:noise} & $[0, \infty)$ \\
${q}^\star_t$       & Ideal capital returns             & Sec. \ref{sec:model_firm} Eq. \eqref{eq:ideal_return} & \\
$\xi$           & Investment risk process           & Sec. \ref{sec:model_risk} Eq. \eqref{eq:real_returns} & $[0, 1]$\\
$\mu_t^q$       & Estimated expected return         & Sec. \ref{sec:model_behaviour} Eq. \eqref{eq:ema_mu} & \\
$\sigma_t^q$    & Estimated investment volatility   & Sec. \ref{sec:model_behaviour} Eq. \eqref{eq:ema_sigma} & \\
$\mathcal{S}_t$ & Estimated Sharpe ratio            & Sec. \ref{sec:model_behaviour} Eq. \eqref{eq:sharpe} & \\
$\mathcal{C}_t$ & Consumption confidence            & Sec. \ref{sec:model_behaviour} Eq. \eqref{eq:cons_confidence} & [-1, 1]\\
$\Sigma_t$      & Unbounded sentiment               & Sec. \ref{sec:model_behaviour} Eq. \eqref{eq:sentiment_level} & \\
\\

\multicolumn{4}{l}{\textbf{Studied Parameters}}\\
$\delta$        & Depreciation rate                 & Sec. \ref{sec:model_household} Eq. \eqref{eq:capital} & $0.001,0.005,0.02$ \\
a               & Investment risk multiplier        & Sec. \ref{sec:model_risk} Eq. \eqref{eq:real_returns_noise} & $15$ \\
$c_0$           & Consumption confidence threshold  & Sec. \ref{sec:model_behaviour} Eq. \eqref{eq:cons_confidence} & $[0,0.025]$ \\
$\lambda$       & Memory kernel for the Sharpe ratio& Sec. \ref{sec:model_behaviour} Eq. \eqref{eq:ema_mu} & $0.607,0.905,0.951,0.98$\\
$\nu$           & Interpolation between $\mathcal{S}_t$ and $\mathcal{C}_t$ & Sec. \ref{sec:model_behaviour} Eq. \eqref{eq:sentiment_level} & $0.75, 1.0$\\
\\

\multicolumn{4}{l}{\textbf{Fixed Parameters}}\\
$\gamma$        & Disutility of labour              & Sec. \ref{sec:model_household} Eq. \eqref{eq:utility_function} & 1 \\
$\pi_t$         & Inflation rate                    & Sec. \ref{sec:model_household} Eq. \eqref{eq:income}  & 0.1\% \\
$r_t$           & Real interest rate                & Sec. \ref{sec:model_household} Eq. \eqref{eq:income} & 0.15\%\\
$p_t$           & Price level                & Sec. \ref{sec:model_household} Eq. \eqref{eq:income} & 1\\
$\alpha$        & Capital share in production       & Sec. \ref{sec:model_firm} Eq. \eqref{eq:prod} & $1/3$ \\
$1/(1+\rho)$    & Elasticity of substitution $k$ vs. $n$ & Sec. \ref{sec:model_firm} Eq. \eqref{eq:prod} & -7 \\
$z_0$ & Baseline value of the total factor productivity &  Sec. \ref{sec:model_firm} Eq. \eqref{eq:noise} & 0.05 \\
$\eta$          & Autocorrelation of total factor productivity     & Sec. \ref{sec:model_firm} Eq. \eqref{eq:noise} & 0.5 \\
$\theta_c$      & Consumption rate transition width       & Sec. \ref{sec:model_behaviour} Eq. \eqref{eq:cons_confidence} & 300 \\
$G_{\min}$        & Minimum consumption rate          & Sec. \ref{sec:model_behaviour} Eq. \eqref{eq:logistic} & 0.05 \\
$G_{\max}$        & Maximum consumption rate          & Sec. \ref{sec:model_behaviour} Eq. \eqref{eq:logistic} & 0.95 \\
$\mathcal{N}$     & Scaling factor for the Sharpe ratio $\mathcal{S}_t$ & Sec. \ref{sec:model_behaviour} Eq. \eqref{eq:sharpe} & 1/4 \\
$F_{\min}$        & Minimum capital allocation           & Sec. \ref{sec:model_behaviour} Eq. \eqref{eq:allocation} & 0.0 \\
$F_{\max}$        & Maximum capital allocation           & Sec. \ref{sec:model_behaviour} Eq. \eqref{eq:allocation} & 1.0 \\
$\theta_k$      & Allocation transition width       & Sec. \ref{sec:model_behaviour} Eq. \eqref{eq:allocation} & 15 \\
\\

\multicolumn{4}{l}{\textbf{Additional Parameters}}\\
$\Xi_k$         & Weighted proportion of time when $k_t < n_t$     & Sec. \ref{sec:phases} Eq. \eqref{eq:xi_k} & $[0, 1]$ \\
$\Xi_c$         & Weighted proportion of time when $c_t < c_0$    & Sec. \ref{sec:phases} Eq. \eqref{eq:xi_c} & $[0, 1]$ \\
$[\mathcal{L}_k,\mathcal{H}_k]$ & Low/High frequency $\Xi_k$ & Sec. \ref{sec:phases} & \\
$[\mathcal{L}_c,\mathcal{H}_c]$ & Low/High frequency $\Xi_c$ & Sec. \ref{sec:phases} & \\
$\mathcal{T}_{\lambda}$ & Timescale of allocation $S_t$ & Sec. \ref{sec:model_dynamics} & $1/|\log(\lambda)|$ \\
$\mathcal{T}_\eta$ & Timescale of TFP shock $z_t$ & Sec. \ref{sec:model_dynamics}  & $1/|\log(\eta)|$ \\
$\mathcal{T}_{\delta}$ & Timescale of capital $k_t$ & Sec. \ref{sec:model_dynamics}  & $1/|\log(1-\delta)|$ \\
$\mathcal{T}_<$ & Average duration of consumption crises & Sec. \ref{sec:mechanisms_memory} &  \\
$\mathcal{T}_>$ & Average duration of high output periods & Sec. \ref{sec:mechanisms_memory} &  \\
\end{tabular}
\end{ruledtabular}
\end{table*}